# Phonon study of rhombohedral BS under high pressure

Kirill A. Cherednichenko,[1,2,3] Petr S. Sokolov,[3] Aleksandr Kalinko,[1,4] Yann Le Godec,[2] Alain Polian,[2] Jean-Paul Itié [1] and Vladimir L. Solozhenko [3,*]

[1] *Synchrotron SOLEIL, 91192 Gif-sur-Yvette, France*

[2] *IMPMC, UPMC Sorbonne Universités, CNRS UMR 7590, 75005 Paris, France*

[3] *LSPM–CNRS, Université Paris Nord, 93430 Villetaneuse, France*

[4] *Institute of Solid State Physics, University of Latvia, LV-1063 Riga, Latvia*

**Abstract**

*Raman spectra of rhombohedral boron monosulfide (r-BS) were measured under pressures up to 34 GPa at room temperature. No pressure-induced structural phase transition was observed, while strong pressure shift of Raman bands towards higher wavenumbers has been revealed. IR spectroscopy as a complementary technique has been used in order to completely describe the phonon modes of r-BS. All experimentally observed bands have been compared with theoretically calculated ones and modes assignment has been performed. r-BS enriched by $^{10}B$ isotope was synthesized, and the effect of boron isotopic substitution on Raman spectra was observed and analyzed.*

## I. Introduction

Boron and its compounds are of great interest due to a rich variety of remarkable properties, which are of both fundamental and technological interest. Among these properties are high hardness and mechanical strength, low density, high melting temperatures, extraordinary chemical and thermal stability, etc., that can be explained by the structures of boron-rich solids consisting of three-dimensional rigid network of $B_{12}$ icosahedra [1]. However, boron compounds are not limited to the $B_{12}$-based structures. Recently new boron sulfides have been synthesized at high pressure i.e. *r*-BS [2], $B_2S_3$ (II) and $B_2S_3$ (III) [3]. $B_2S_3$ (III) has a tetragonal structure (space group $I4_1/a$) based on $BS_4$-tetrahedra, while *r*-BS is a layered compound with γ-GaS structure (space group R3m, $C_{3v}$ point group). *r*-BS was revealed to be layered semiconductor (estimated band gap value is 3.4 eV [2]), one layer being built by B–B pairs aligned along the c-axis, placed between hexagonal layers of S atoms (Fig. 1), rotated by π/3 relative to each other. The bonding scheme is hence the same as that of the $A^{III}B^{VI}$ layered compounds, *i.e.* strong covalent intralayer bonds and weak Van der Waals interlayer bonds. Hence, the physical properties of *r*-BS should be affected by its anisotropic structure. Unlike other $A^{III}B^{VI}$ semiconductors (GaS, GaSe, InSe, etc.) [4-9], the properties of *r*-BS are very ill-studied, especially, under pressure.

Raman and IR spectroscopy is powerful tools to study vibrational properties of solids and their modifications due to structural defects isotopic composition, which is of particular importance in the case of boron compounds [10-12]. In this paper we report Raman and IR studies of *r*-BS at ambient and high (up to 34 GPa) pressure. Experimental results have been compared with calculated phonon properties of *r*-BS, and bands assignment has been implemented.

## II. Experimental technique

Polycrystalline *r*-BS was synthesized at 7.5 GPa and 2200 K from amorphous boron (Johnson Matthey, 99%) and sulfur (Johnson Matthey, spectrographic grade) powders mixed in the 1:1 molar ratio using a toroid-type high-pressure apparatus with specially designed high-temperature cell [13]. Boron nitride (grade AX05, Saint-Gobain) capsules were used to isolate the reaction mixture from the graphite heater. Isotope enriched *r*-$^{10}$BS was synthesized by the same method from amorphous $^{10}$B (enrichment > 95%) (for convenience, further we will mark the *r*-BS with natural B isotope distribution as *r*-BS and the phase enriched by $^{10}$B isotope as *r*-$^{10}$BS). The structure and phase purity of as-synthesized yellow-green polycrystalline samples were confirmed by powder X-ray diffraction (G3000 TEXT Inel diffractometer, Cu K$\alpha$1 radiation). The lattice parameters of both *r*-BS and *r*-$^{10}$BS are close to the literature data ($a = 3.0522$ Å, $c = 20.4119$ Å) [2] and presented in Table I. Bond lengths and angles of *r*-BS synthesized in the present work, as well as literature values [2], are listed in Table I.

*In-situ* high-pressure experiments were carried out in a membrane diamond anvil cell (DAC) [14] with 250 µm culet anvils. A small (~20 µm) sample of *r*-BS was loaded into a 90-µm hole drilled in a rhenium gasket pre-indented down to 35 µm. Neon pressure transmitting medium has been used. Pressure in the DAC was determined by the ruby luminescence technique [15] using the calibration by Mao et al. [16]. Pressure was measured before and after each run; drift during experiment was less than 0.4 GPa in the 2–34 GPa pressure range.

Raman spectra were excited with the 632.8 nm line of a He-Ne laser (10 µm beam spot) and recorded in the 200-2000 cm$^{-1}$ range using Horiba Jobin Yvon HR800 Raman spectrometer. The spectrometer was calibrated using a single crystal of cubic Si at room temperature. A laser power at the sample was estimated to be less than 30 mW. No effect due to laser heating of the sample was observed. The positions of individual and overlapped Raman bands have been evaluated by fitting the experimental profile to the Pearson shape function using DatLab software [17]. Raman spectra of *r*-BS were measured in 23 pressure points from 1.8 to 34.0 GPa at room temperature. The Raman study of *r*-$^{10}$BS was performed only at ambient conditions with the aim to identify the modes impacted by the isotopic effect, and thus to help understand the modes symmetry.

The Fourier transform infrared (FTIR) absorption spectra in mid-infrared range (450-4000 cm$^{-1}$) were recorded using a Bruker IFS 125HR spectrometer. Samples were uniformly mixed with KBr powder and pressed into pellets.

## III. Calculation details

In the present work the *r*-BS has been studied using first principles LCAO calculations performed using the CRYSTAL09 code [18]. For the boron atoms, we used the all-electron basis sets, which were optimized in earlier calculations [19]. The core electrons of the sulfur atom were excluded from consideration using the effective small-core pseudopotential (ECP) and the corresponding atomic basis set [20], which excludes any diffuse Gaussian-type orbitals with the exponents less than 0.1. In the CRYSTAL09 code, the accuracy in evaluating the Coulomb series and the exchange series is controlled by a set of tolerances, which were taken to be ($10^{-8}$, $10^{-8}$, $10^{-8}$, $10^{-8}$, and $10^{-16}$). The Monkhorst–Pack scheme [21] for a 8×8×8 k-point mesh in the Brillouin zone was applied. Self-consistent field calculations were performed for hybrid DFT–HF WCGGA–PBE-16% functional [22]. The percentages 16% define the Hartree–Fock admixture in the exchange part of DFT functional.

The full structure optimization procedure according to the energy minima criterion has been performed for *r*-BS. The optimized structural parameters, such as the lattice parameters ($a$, $b$, $c$), unit cell volume ($V_0$), bond lengths and angles (see Fig. 1) are compared with experimental ones and listed in Table I.

The bulk modulus and its first derivative for *r*-BS have been calculated using routine implemented in the CRYSTAL09 code. In order to obtain E(V) dependence, the unit cell volume was varied from 76% to 100% of the equilibrium volume ($V_0$). At each volume the structure optimization was performed. Obtained E(V) dependence has been fitted to the Murnaghan equation of state [23] taking $V_0 = 57.61$ Å$^3$ and $E_0 = -69.97$ eV. P(V) dependence was estimated using equations (1), (2).

$$E(V) = E_0 + V_0 B_0 \left\{ \frac{1}{B_0'(B_0'-1)} \left(\frac{V}{V_0}\right)^{1-B_0'} + \frac{V}{V_0 B_0'} - \frac{1}{B_0'-1} \right\}; \qquad (1)$$

$$P(V) = \frac{B_0}{B_0'} \left[ \left(\frac{V}{V_0}\right)^{B_0'} - 1 \right] \qquad (2)$$

The phonon frequencies for *r*-$^{11}$B$_{(100\%)}$S and *r*-$^{10}$B$_{(100\%)}$S (further they will be marked as *r*-$^{11}$BS* and *r*-$^{10}$BS*, respectively) have been calculated using the direct (frozen-phonon) method [18,24]. The calculated values are reported in Table II. The phonon frequencies at selected pressure points up to 34 GPa have been calculated using optimized geometries for corresponding reduced volume unit cells.

### IV. Results and discussion

According to the powder X-ray diffraction (XRD) data [2], *r*-BS has a space group R3m (160), the same as γ-GaS [4] and γ-GaSe [6]. Rietveld refinement was not performed in our study because boron is a low-Z element. Thus, the refined atomic coordinates determined in [2] have been used in combination with lattice parameters obtained from our XRD data. As one can see from Fig. 1 and Table I, the *r*-BS trigonal anti-prism [2] is slightly distorted. Accurate measurements of all bond lengths and angles in *r*-BS and *r*-$^{10}$BS obtained from experimental XRD data reveal the same distortion. However, the geometry optimization procedure in the LCAO calculations for *r*-BS* doesn't support the experimentally observed distortion. This fact could not be explained yet as well as slightly different *c* parameters of *r*-$^{10}$BS and *r*-BS.

*r*-BS (or γ-BS) has one layer in the primitive rhombohedral cell (rhombohedral setting). Structural unit of *r*-BS is trigonal anti-prism containing B-B pairs and sulfur atoms in vertices, forming two hexagonal sheets of S rotated around the c-axis by π/3 relative to each other. Consequently, the coordination number of boron atoms in *r*-BS is four. *r*-BS can be represented in a hexagonal setting with three layer in a crystallographic cell (Fig. 1). It can be compared with the MX$_2$ transition metal chalcogenides where the B-B pair is replaced by an M atom.

In layered crystals, the vibration modes may be separated in the modes of the isolated layer (internal modes in molecular crystals), and the whole crystal modes. Since the layers contain the same number of atoms and due to weak Van der Waals interlayer interactions, the number of modes from isolated layer should be multiplied by the number of layers in the primitive unit cell. That gives rise to "rigid layer modes" (similar to the external modes in molecular crystals) and "Davydov multiplets". However, the Raman studies on the GaSe polytypes [25-30] have proved that the layer modes are not dependent on the number of layers.

*r*-BS has 4 atoms in the unit cell of rhombohedral crystal structure. Thus, 12 normal modes of vibration at the zone center are described by the irreducible representation of the C$_{3v}$ point group:

$$\Gamma = 4A_1 + 4E \qquad (3)$$

where the E modes are doubly degenerate. All the optic modes are both infrared and Raman active and thus there should be 6 non-degenerate Raman active modes, since $A_1(1)$ and $E(1)$ modes (Fig. 2(a)) are the acoustic ones. An irreducible representation is assigned to a set of atomic displacement pattern in Fig. 3, in a manner similar to that previously used [27]. In this Figure the bracketed numbers associated with the group representations have no physical meaning and are used only as labels.

At ambient conditions Raman spectra were investigated in the 200–2500 cm$^{-1}$ frequency range. Five Raman active modes of $r$-BS have been observed in the 200–1200 cm$^{-1}$ range (Fig. 2(a)). In the case of isotope enriched $r$-$^{10}$BS, only four bands have been observed in the same frequency domain. No peak was observed above 1200 cm$^{-1}$ for both compounds. In the Raman spectra of $r$-BS and $r$-$^{10}$BS less than six bands predicted from the irreducible representation of non-degenerate Raman modes were observed. The frequencies of all modes for $r$-$^{10}$BS and $r$-BS are presented in Table II.

IR spectrum of $r$-BS (Fig. 2(b)) in the 200–4000 cm$^{-1}$ frequency range shows many bands. Most of them can be explained by the presence of impurities and adsorbed water. However, based on the shape and relative intensities of IR spectrum of one of GaS polymorphs [5] we characterize two bands, which we suppose are vibrational modes of $r$-BS. Our suggestion is supported by the LCAO calculations (see Table II). In the IR spectrum the $r$-BS bands are placed at 656.6 cm$^{-1}$ and 702.9 cm$^{-1}$. As far as in transmission IR spectroscopy only TO modes could be measured, we observed $E(2)$ (TO) and $A_1(2)$ (TO).

The results of our LCAO calculations corresponding to the lowest temperature limit (T = 0 K) agree with the experimentally observed bands wavenumbers (Table II). Based on LCAO calculations performed for $r$-$^{11}$BS* and $r$-$^{10}$BS*, the Raman- and IR-active modes assignment has been performed. The deviation of calculated Raman modes from experimentally observed ones does not exceed 3.4% and 5.6% for $r$-BS and $r$-$^{10}$BS, respectively. The deviation of calculated IR modes from experimentally observed bands for $r$-BS is less than 5.5%. To confirm the obtained results we provided test calculation of phonon frequencies of γ-GaSe. Results of these calculations are presented in Table II. One can see that the sequence of the modes found in Raman spectra (E(3), A1(3), E(4), A1(4)) of γ-GaSe [27] is the same as for $r$-BS. This fact makes it clear that compounds of the same symmetry and almost similar structure exhibit the common sequence of the phonons presented in Raman and IR spectra.

The LCAO calculations have shown that the IR intensity of the E(2) and $A_1(2)$ modes is much higher than that of the other modes. In spite of the fact that all modes are both Raman and IR active, the most intense IR modes have not been observed in Raman spectra. Taking into account this fact, we suppose that those modes, which have not been observed in Raman spectra are observed in IR spectra.

The isothermal bulk modulus of $r$-BS $B_0$ = 37.9 GPa and its first derivative $B_0` $ = 7.5 have been estimated using the E(V) dependence obtained from LCAO calculations (Fig. 4). Volume dependence V(P) fitted to the Murnaghan equation of state is shown in Fig. 5. The rather high value of $B_0´$ is evidently due to the anisotropy of the $r$-BS structure similar to graphite ($B_0´$ = 8.9) [31], hexagonal graphite-like boron nitride ($B_0´$ = 5.6) [32] and turbostratic BN ($B_0´$ = 11.4) [33]. The band gap value $E_g$ = 3.45 eV found from LCAO calculations is in good agreement with the estimation made in Ref. 2 (~3.4 eV). Figure 6 presents the dependence of the lattice parameters on the pressure. The most significant compression occurs along the $c$ axis in contrast to directions along $a$ and $b$ axes, which is consistent with the fact that the interlayer interactions are of the Van der Waals type, whereas the intralayer ones are iono-covalent. Hence, the compressibility along $c$ is much larger than in other directions and results in the rather low value of $B_0$. Compression of the B-B and B-S bonds in the layer are similar (Fig. 7) similar to that for GaSe [7,8] and GaSe [28,30].

There are significant differences between Raman spectra of *r*-BS with natural B isotope distribution (80% $^{11}$B – 20% $^{10}$B) and of *r*-BS enriched in $^{10}$B isotope (95% $^{10}$B – 5% $^{11}$B). As expected, all the bands of the $^{10}$B-enriched sample are shifted to high-frequencies. Isotope shift exceeds 20 cm$^{-1}$ for the pair of high-frequency phonons while it does not exceed 2 cm$^{-1}$ for the low-frequency ones. It has been concluded that frequencies of E(4) and A$_1$(4) are significantly influenced by the oscillation frequencies of the boron atoms, and therefore by boron atoms mass change. This conclusion has been confirmed by LCAO calculations i.e. experimentally observed bands of *r*-$^{10}$BS and theoretically predicted phonons revealed good coincidence (see Fig. 2(a)).

The rather intense band at 1037.9 cm$^{-1}$ observed in Raman spectrum of *r*-BS is not predicted by LCAO calculations. It has been observed only in the Raman spectra of *r*-BS with the natural boron isotope content. The coincidence of XRD patterns of *r*-$^{10}$BS and *r*-BS excludes any possibility of structural difference or systematic alternating defects. Moreover, it has been noted [27] that the vibrational frequencies are dominated by the intralayer forces and that changes in stacking sequence has no influence on the frequency of the "internal" layer modes. However, invisible for XRD technique random structural defects may occur in the polycrystalline sample and thus could lead to unpredictable bands arrangement and relative intensities or even appearance of band splitting in Raman and IR spectra. Moreover, one can remark that Raman spectra of GaSe and GaS polymorphs (structural analogs of *r*-BS) have the same doublet at high-frequency wavenumbers [27,29]. Hoff *et al.* [27] showed that some of the bands can appear as result of combination of A(LO) – E(LO) and A(TO) – E(TO) modes. This could be an explanation of the doublet appearance, but proving this assumption requires techniques that are not available in the present work.

The pressure dependence of the five vibrational modes has been measured in *r*-BS up to 34 GPa. Raman spectra at different pressures are shown in Fig. 8, indicating rather strong phonon's shift. During the compression, all the lines have shifted monotonically toward high frequencies (Fig. 9), and no new line appeared (Fig. 8). Thus, one can conclude that no phase transition nor amorphization is observed in the studied pressure range. In contrast, one should note that GaS and GaSe do have phase transitions at 18÷19 GPa [28] and 24÷25 GPa [8], respectively. Phase transition of β-GaS observed in [30] was followed by total disappearing of all Raman bands except one. It has been stated that at ambient pressure intralayer bonds are one order of magnitude larger than interlayer restoring forces [30]. Thus, an increase of the latter under pressure brings them into the range of the intralayer forces and, therefore, structure becomes quasi-three-dimensional near the phase transition. A broad band in the 500 – 625 cm$^{-1}$ range is observed, but no correlation between band's intensity, frequency shift and pressure applied to the system could be found. Thereby we suppose that the origin of this band is connected with the non-homogeneity of the powder sample under study. There are strong grounds to believe that no new additional bands appear up to the highest experimental pressure due to the provided LCAO calculations. One can see from Fig. 9 that the theoretical calculations completely match with Raman bands shift observed experimentally in the whole studied pressure range.

Simultaneous growing of the background and bands broadening has been observed. High-frequency bands at 1037.9 and 1046.6 cm$^{-1}$ became indistinguishable above 29 GPa. It should be noted that there are monotonic dependencies of full width at half maximum (FWHM) of the second (~320 cm$^{-1}$) and third (~700 cm$^{-1}$) bands with increasing pressure. That of the first band can be approximated by a straight line. The absence of sharp breaks in the pressure dependencies of the FWHM may be considered as indirect evidence of absence of phase transition in *r*-BS over the investigated pressure range.

In order to estimate the Grüneisen parameters of observed Raman modes we fitted the Raman modes pressure dependencies (Fig. 9) by the same equation as in Ref. 34:

$$\omega = \omega_0 \cdot \left(1 + P \cdot \frac{\delta_0}{\delta'}\right)^{\delta'}, \text{ where } \delta_0 = \left(\frac{d\ln\omega}{dP}\right)_{P=0} \text{ and } \delta' = \delta_0^2 \left(\frac{d^2\ln\omega}{dP^2}\right)_{P=0}. \quad (4)$$

A least-squares fit of Eq. (4) to the experimental data yields the values of first-order parameters ($\delta_0$) 0.0064(5) GPa$^{-1}$ for E(3) mode (209.4 cm$^{-1}$), 0.0136(3) GPa$^{-1}$ for A$_1$(3) mode (319.3 cm$^{-1}$), 0.0071(2) GPa$^{-1}$ for E(4) mode (686.5 cm$^{-1}$), and 0.0048(1) GPa$^{-1}$ for A$_1$(4) mode (1046.6 cm$^{-1}$). Also the pressure dependence of the band at 1037.9 cm$^{-1}$ has been fitted by Eq. (4), and $\delta_0$ value of 0.0048(2) GPa$^{-1}$ has been determined. The Grüneisen parameters $\gamma_G = B_0 \delta_0$ of all bands are presented in Table II.

All the observed phonons moved to high frequency, but the extension of these shifts was different (Fig. 9). The dashed lines have been fitted using the quadratic equation:

$$\omega = \omega_0 + \omega_1 \cdot P + \omega_2 \cdot P^2 \qquad (5)$$

The parameters $\omega_1$ used in Eq. (5) are given in Fig. 9. Observed coefficients $\omega_1$ and $\omega_2$ have values typical for boron-rich compounds [10-12]: $\omega_1$ values are not exceeding 5, while $\omega_2$ values are slightly negative tending to zero. Besides, these coefficients are of the same order of magnitude as those of the corresponding modes of GaSe [7] and GaS [30].

The initial slope of the low-frequency mode E(3) is much smaller than that of the A$_1$(3), E(4) and A$_1$(4) modes. Also (as shown in Fig. 3) only E(3) reveals the same amplitude of vibration for boron and sulfur atoms. Summing up, we can conclude the particularity of the E(3) mode with respect to the other observed modes. This mode is similar to the E mode (60 cm$^{-1}$) of γ-GaSe, E" mode (60 cm$^{-1}$) in ε-GaSe and E$_{1g}$ mode (74 cm$^{-1}$) in β-GaS, because all these modes have the same displacement pattern (rigid half layer shear modes). Moreover, in these compounds, these modes have low pressure coefficients, in contrast to the other modes. In Ref. 30, the peculiar pressure coefficient of the E$_{1g}$ mode was explained by the similarity of its displacement pattern with an edge-of-the-zone TA mode as regards the destabilizing effect of pressure upon electronic contributions to restoring forces. The pressure coefficient of the E(3) can be explained in the same way. It is a half-layer shear mode, involving mainly B-B restoring forces, as far as vibration amplitudes for B and S atoms are the same in this mode. Since this mode does not change much the B-S distance within the same half-layer, this vibration will be viewed as a transverse acoustical (TA) mode of BS molecules, on a chain along the *c* axis. Edge-of-the-zone TA modes in 3D crystals are known to exhibit negative coefficients of the pressure dependence. Under pressure the electronic charges involved in the first neighbors bonds delocalize towards interlayer space, which soften the total restoring spring, and results in a negative pressure coefficient. We assumed that explanation given in [30] can be used also for E(3) mode. Low coefficient of the pressure dependence of E(3) mode is a result of two contributions: a positive one coming from the increase with pressure of the interlayer interaction and therefore distance decrease between the atoms, and a negative one coming from the shear motion and the TA-like character in this mode. The total may be slightly positive, zero, or slightly negative i.e. much less than for other intralayer modes.

The same explanation concerning weak pressure dependence of the low-frequency Raman band was proposed in [7]. Very small pressure coefficient of such mode can be explained by the compensation of the decrease in the B-B bond length, which leads to the frequency increase under pressure, and the charge delocalization, which tends to depopulate the B$_2$ radical environment and thus decrease the B-B bond strength. The same effect can be observed in diatomic molecules under the pressure (I$_2$, H$_2$ *etc.*), where the charge transfer occurs between the space towards to the intermolecular region, tending to symmetrization of the bonds and eventually leading to the same intensity of intra- and intermolecular ones. In other words, the charge density between the layers increases with the pressure because of interlayer space decrease and also of the total effective number of electrons squeezed out from intralayer space.

## V. Conclusions

Rhombohedral boron monosulfide, *r*-BS has been studied by Raman and IR spectroscopy at ambient and high pressure. Basing on the results of LCAO calculations, observed isotope effect and Raman spectra of isostructural $A^{III}B^{VI}$ compounds, the correlation between Raman and IR bands and phonon modes was performed. The evolution of Raman bands of *r*-BS has been studied up to 34 GPa at ambient temperature. Modes behavior under pressure was described and explained. Finally, it was found that there is no phase transition of *r*-BS in the explored pressure range.

## Acknowledgements

The authors thank Dr. Pascal Munsch (IMPMC) for help in DAC preparation, Dr. Alexandre Tallaire (LSPM–CNRS) for assistance in Raman measurements and Dr. Oleksandr O. Kurakevych (IMPMC) for stimulating discussions. This work was financially supported by the Agence Nationale de la Recherche (grant ANR-2011-BS08-018).

## References

[1]. B. Albert, H. Hillebrecht, *Angew. Chem. Int. Ed.* **2009,** *48*, 8640.

[2]. T. Sasaki, H. Takizawa, K. Uheda, T. Endo, *Phys. Status Solidi B* **2001**, *223*, 29.

[3]. T. Sasaki, H. Takizawa, K. Uheda, T. Yamashita, T. Endo, *J. Solid State Chem.* **2002**, *166*, 164.

[4]. M.P. Pardo, J. Flahaut, *Mater. Res. Bull.* **1987**, *22*, 323.

[5]. M.J. Taylor, *J. Raman Spectrosc.* **1973**, *1*, 355.

[6]. K. Schubert, E. Doerre; M. Kluge, *Z. Metallkd.* **1955**, *46*, 216.

[7]. M. Gauthier, A. Polian, J.M. Besson, A. Chevy, *Phys. Rev. B* **1989**, *40*, 3837.

[8]. J. Pellicer-Porres, A. Segura, C. Ferrer, V. Munoz, A. San Miguel, A. Polian, J. P. Itié, M. Gauthier, S. Pascarelli, *Phys. Rev. B*, **2002**, *65*, 174103.

[9]. D. Olguin, A. Cantarero, C. Ulrich, K. Syassen, *Phys. Stat. Sol. B*, **2003**, *235*, 456.

[10]. V.L. Solozhenko, O.O. Kurakevych, P. Bouvier, *J. Raman Spectrosc.*, **2009**, *40*, 1078.

[11]. S. Ovsyannikov, A. Polian, P. Munsch, J.-C. Chervin, G. Le Marchand, T.L. Aselage, *Phys. Rev. B* **2010**, *81*, 140103.

[12]. J.W. Pomeroy, M. Kuball, H. Hubel. N.W.A. van Uden, D.J. Dunstan, R. Nagarajan, and J.H. Edgar, *J. Appl. Phys.* **2004**, *96*, 910.

[13]. V.A. Mukhanov, P.S. Sokolov, V.L. Solozhenko, *J. Superhard Mater.*, **2012**, *34*, 211.

[14]. J. C. Chervin, B. Canny, J. M. Besson, Ph. Pruzan, *Rev. Sci. Instrum.*, **1995**, *66*, 2595.

[15]. G. J. Piermarini, S. Block, *Rev. Sci. Instrum.* **1975**, *46*, 973.

[16]. H. K. Mao, J. Xu, P. M. Bell, *J. Geophys. Res.* **1986**, *91*, 4673.

[17]. K. Syassen, Computer Code DATLAB, Max Planck Institute, Stuttgart, Germany (**2012**).

[18]. R. Dovesi, V.R. Saunders, C. Roetti, R. Orlando, C.M. Zicovich-Wilson, F. Pascale, et al., *CRYSTAL09 user's manual*. Torino: University of Torino (**2009**).

[19]. R. Orlando, R. Dovesi, C. Roetti, *J Phys.-Cond. Mat.*, **1990**, *2*, 7769.

[20]. A. Bergner, M. Dolg, W. Kuechle, H. Stoll, H. Preuss, *Mol. Phys.*, **1993**, *80*, 1431.

[21]. H.J. Monkhorst, J.D. Pack, *Phys Rev B*, **1976**, *13*, 5188.

[22]. Z. Wu, R.E. Cohen, *Phys Rev B*, **2006**, *73*, 235116.

[23]. Murnaghan, *Proc. Natl. Acad. Sci.*, **1944**, *30*, 244.

[24]. F. Pascale, C.M. Zicovich-Wilson, F. Lopez, B. Civalleri, R. Orlando, R. Dovesi, *J. Comput. Chem.*, **2004**, *25*, 888.

[25]. A. Polian, K. Kunc, A. Kuhn, *Solid State Comm.*. **1976**, *19*, 1079.

[26]. A. Kuhn, A. Chevy, R. Chevalier, *Phys. Status Solidi* A **1975**, *31*, 469.

[27]. R.M. Hoff, J.C. Irwin, R.M.A. Lieth, *Can. J. Phys*. **1975**, *53*, 1606.

[28]. A. Polian, J.M. Besson, M. Grimsditch, H. Vogt, *Phys. Rev. B* **1982**, *25*, 2767.

[29]. J.C. Irwin, R.M. Hoff, B.P. Clayman and R.A. Bromley, *Solid State Comm.*, **1973**, *13*, 1531.

[30]. A. Polian, J.M. Besson, J.C. Chervin, *Phys. Rev. B* **1980**, *22*, 3049.

[31]. M. Hanfland, H. Beister, K. Syassen, *Phys. Rev. B* **1989**, *39*, 12598.

[32]. V.L. Solozhenko, G. Will, F. Elf, *Solid State Comm.*, **1995**, *96*, l.

[33]. V.L. Solozhenko, E.G. Solozhenko, *High Pressure Res.*, **2001**, *21*, 115.

[34]. V.L. Solozhenko, O.O. Kurakevych, Y. Le Godec, A.V. Kurnosov, A.R. Oganov, *J. App. Phys.*, **2014**, *116*, 033501.

Table I. Experimental and calculated structural parameters of *r*-BS, *r*-$^{10}$BS and *r*-BS* (LCAO) (hexagonal setting).

| Parameter | *r*-BS | *r*-$^{10}$BS | *r*-BS* (LCAO) | *r*-BS [2] |
|---|---|---|---|---|
| *a*=*b*, Å | 3.054(1) | 3.055(5) | 3.114(7) | 3.052(2) |
| *c*, Å | 20.482(2) | 20.348(4) | 20.749(7) | 20.411(9) |
| $V_0$, Å$^3$ | 165.453 | 164.523 | 174.326 | 164.679 |
| B1-B2, Å | 1.700 | 1.689 | 1.687 | 1.694 |
| B1-S1, Å | 1.993 | 1.990 | 1.987 | 1.990 |
| B2-S2, Å | 1.913 | 1.912 | 1.987 | 1.911 |
| S1-S2, Å (intralayer) | 3.384 | 3.359 | 3.366 | 3.372 |
| S1-S2, Å (interlayer) | 3.433 | 3.433 | 3.535 | 3.433 |
| S1-B1-S1, ° | 100.041 | 100.244 | 103.223 | 100.125 |
| S2-B2-S2, ° | 105.918 | 106.075 | 103.233 | 105.983 |
| S1-B1-B2, ° | 117.771 | 117.609 | 115.170 | 117.704 |
| S2-B2-B1, ° | 112.826 | 112.685 | 115.160 | 112.768 |

Table II. Experimental ($\omega_0$) and calculated ($\omega_t$) phonon frequencies of *r*-BS, *r*-$^{10}$BS and $\gamma$-GaSe; *r*-BS isotope shifts [$\Delta\omega = \omega(r$-$^{10}$BS$) - \omega(r$-BS$)$] for the experimentally observed Raman bands and theoretically calculated modes; modes assignment and Grüneisen parameters ($\gamma_G$).

| Modes | Raman spectroscopy | | | | | | | | $\gamma_G$ |
|---|---|---|---|---|---|---|---|---|---|
| | Wavenumber (cm$^{-1}$) | | | | | | | | |
| | Exp [27] | LCAO | Experiment* | | | LCAO** | | | |
| | $\omega_0(\gamma$-GaSe) | $\omega_t(\gamma$-GaSe) | $\omega_0(r$-BS) | $\omega_0(r$-$^{10}$BS) | $\Delta\omega$ | $\omega_t(r$-$^{11}$BS*) | $\omega_t(r$-$^{10}$BS*) | $\Delta\omega$ | |
| E(3) | 59.4 | 62 | 209.4 | 210.9 | 1.5 | 215.3 | 216.8 | 1.5 | 0.243 |
| A$_1$(3) | 135 | 143 | 319.3 | 320.7 | 1.4 | 313.7 | 314.1 | 0.4 | 0.515 |
| E(4) | 211 | 236 | 686.5 | 708 | 21.5 | 671.7 | 699.8 | 28.1 | 0.269 |
| ? | | | 1037.9 | | | | | | 0.182 |
| A$_1$(4) | 309.5 | 338 | 1046.6 | 1071.7 | 25.1 | 1083.2 | 1135.3 | 52.1 | 0.182 |

| Modes | IR spectroscopy | |
|---|---|---|
| | Wavenumber (cm$^{-1}$) | |
| | Experiment | LCAO* |
| | $\omega_0$ (*r*-BS) | $\omega_t$ (*r*-BS) |
| E(2) | 656.6 | 622.5 |
| A$_1$(2) | 702.9 | 672.2 |

* - Isotope contents of *r*-BS and *r*-$^{10}$BS used in experiments are: 20% $^{10}$B-80% $^{11}$B and 95% $^{11}$B-5% $^{11}$B, respectively.

** - LCAO calculations have been performed for (100% $^{11}$B) *r*-BS and (100% $^{10}$B) *r*-BS.

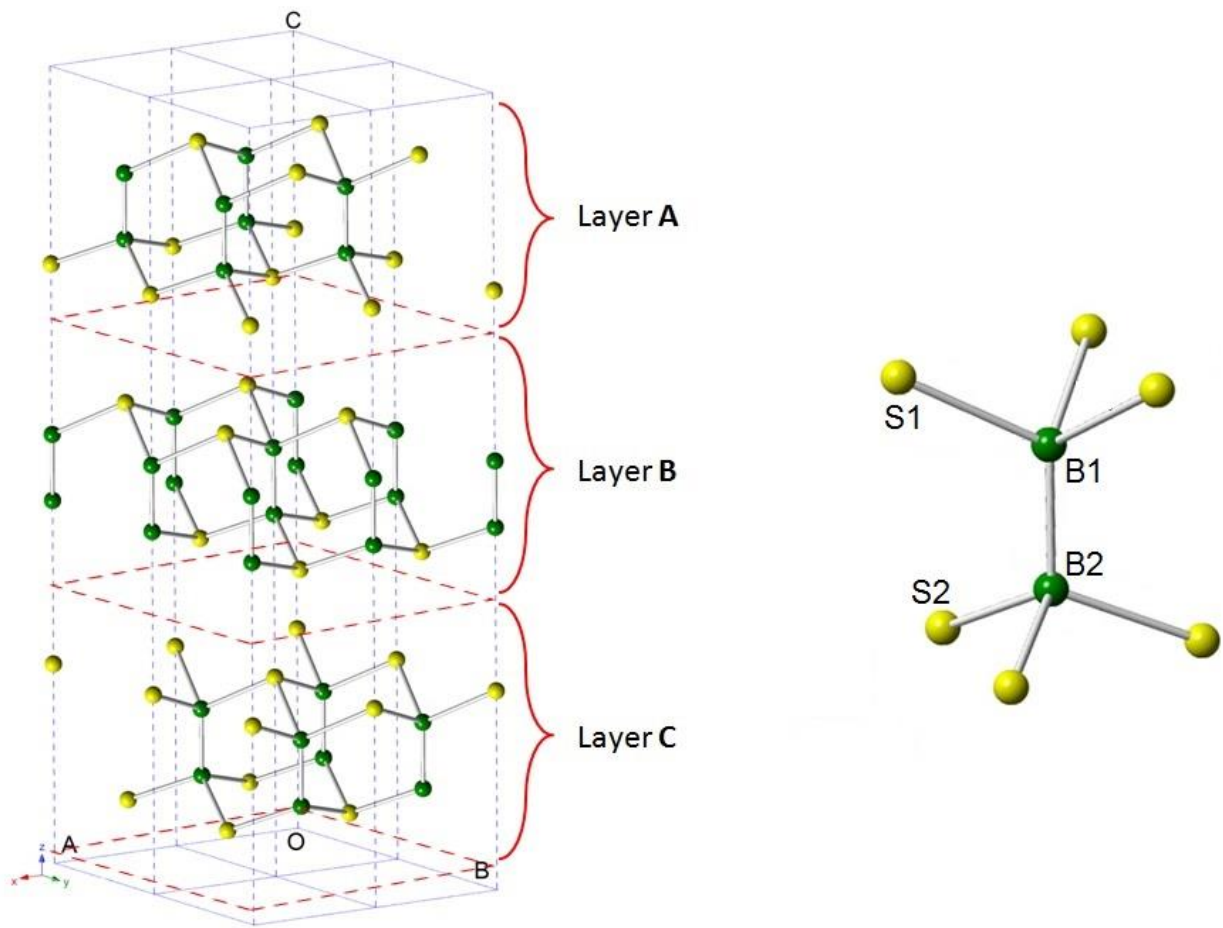

FIG. 1. Representation of the layers stacking and trigonal anti-prism of *r*-BS.

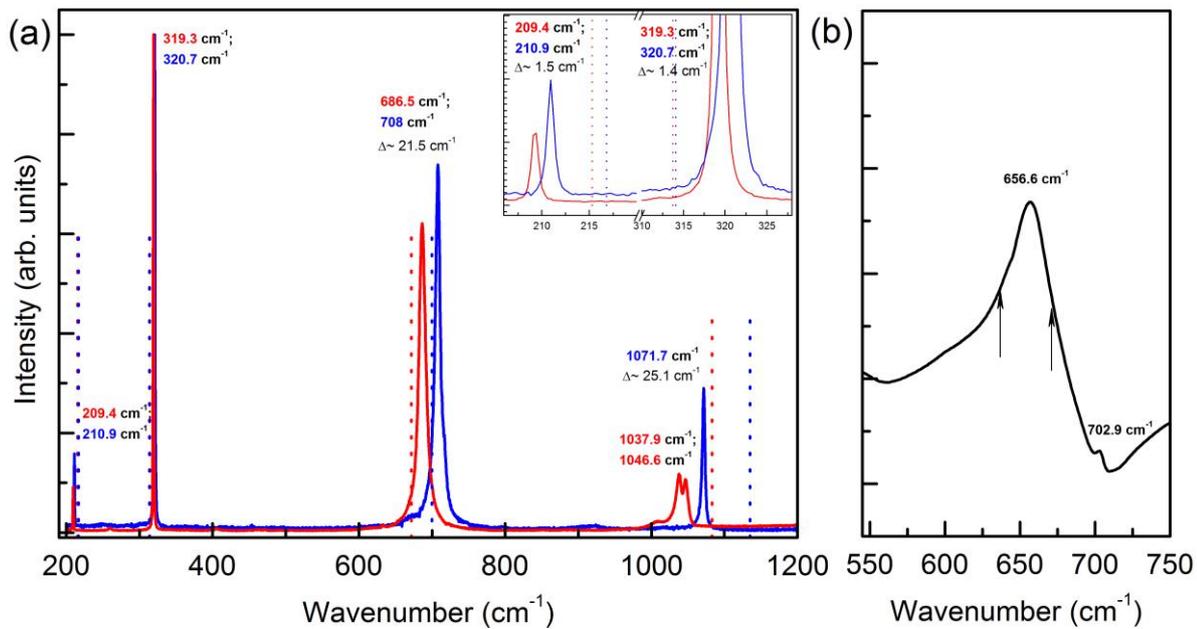

FIG. 2. (a) Experimentally observed (solid lines) and LCAO calculated (dashed lines) Raman bands in *r*-BS (red) and *r*-$^{10}$BS (blue) at ambient conditions. Inset: magnification of the 200-330 cm$^{-1}$ region. Frequency values of experimentally observed modes are indicated.
(b) Experimentally observed and LCAO calculated (arrows) bands in IR spectrum of *r*-BS at ambient conditions.

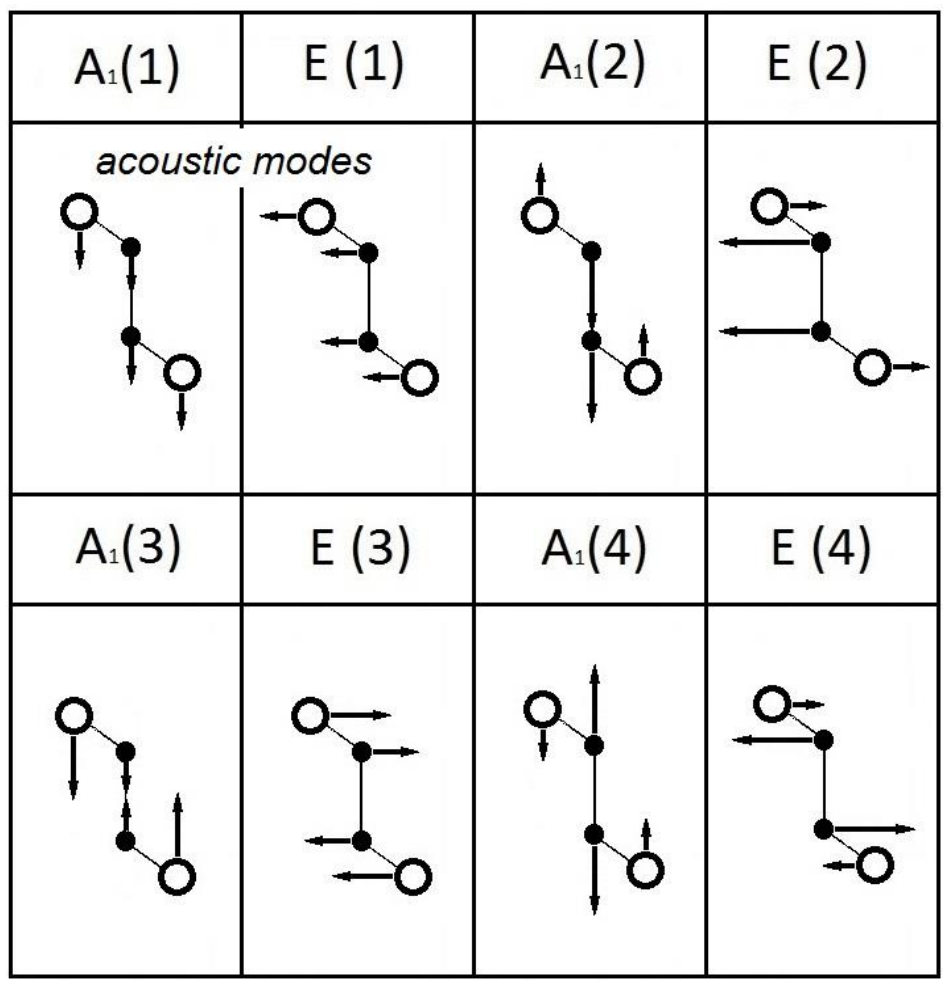

FIG. 3. Normal vibration modes of *r*-BS. The arrows length reflects relative amplitudes of B and S atoms oscillations in the mode. The amplitudes have been estimated from LCAO calculations.

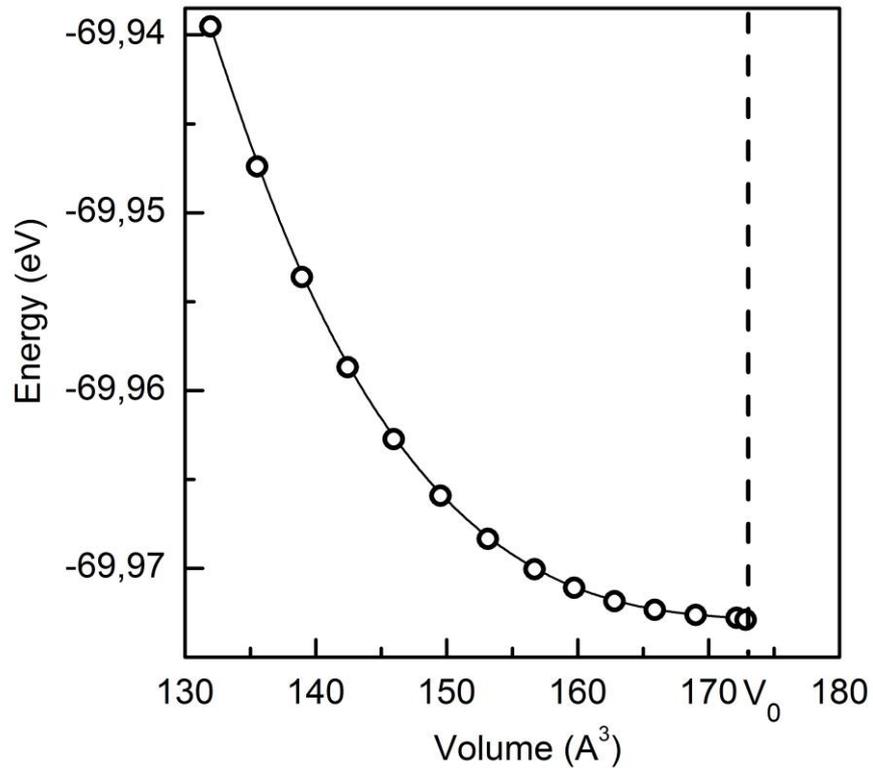

FIG. 4. Energy variation of the *r*–BS unit cell *versus* volume (Murnaghan fit).

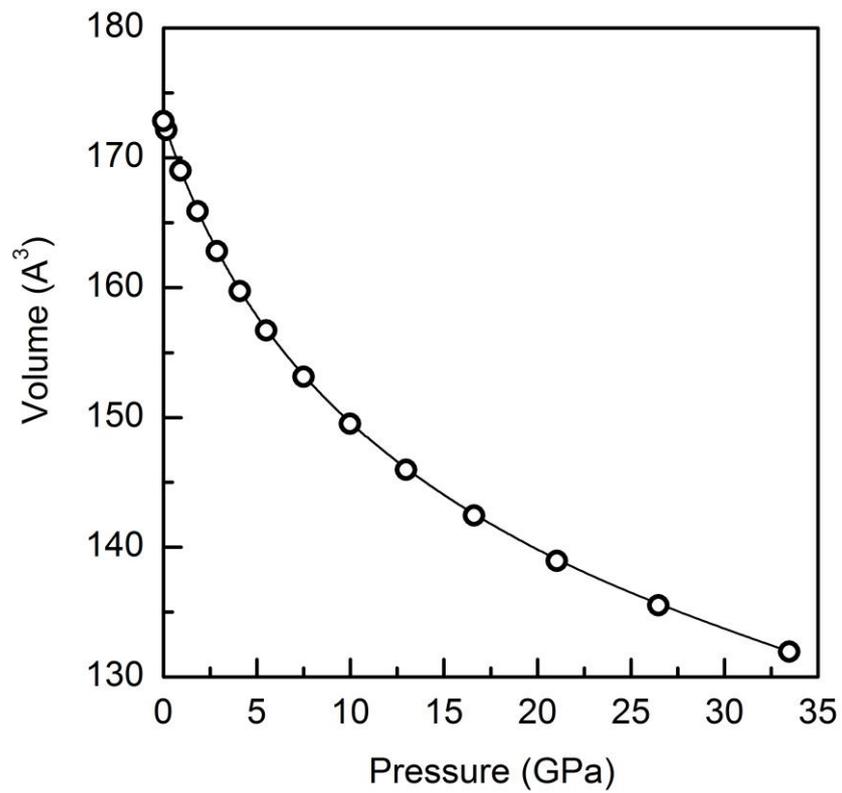

FIG. 5. Calculated volume of the *r*–BS unit cell *versus* pressure (Murnaghan fit).

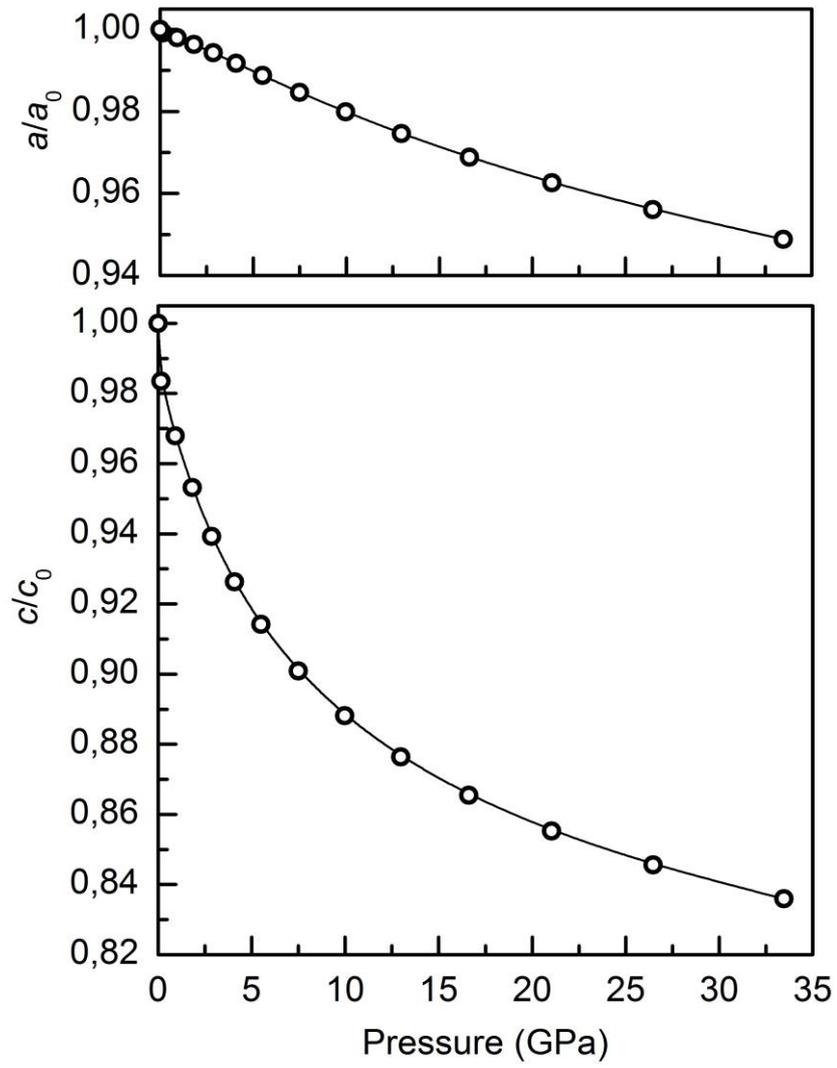

FIG. 6. Relative lattice parameters of *r*-BS *versus* pressure.

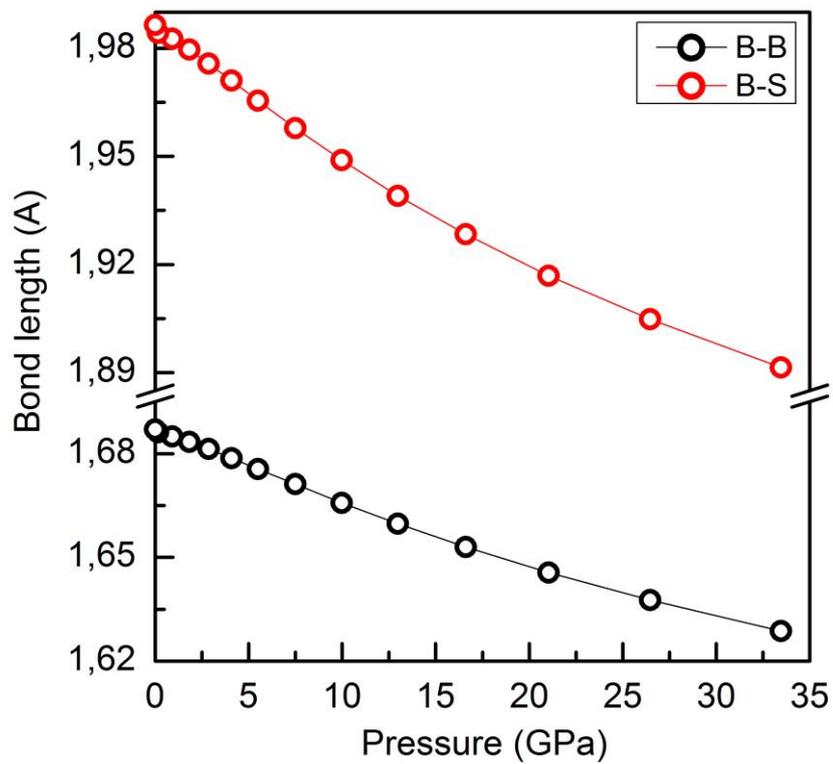

FIG. 7. Calculated B-B and B-S bond lengths in the *r*-BS structural unit (see Fig. 1) *versus* pressure.

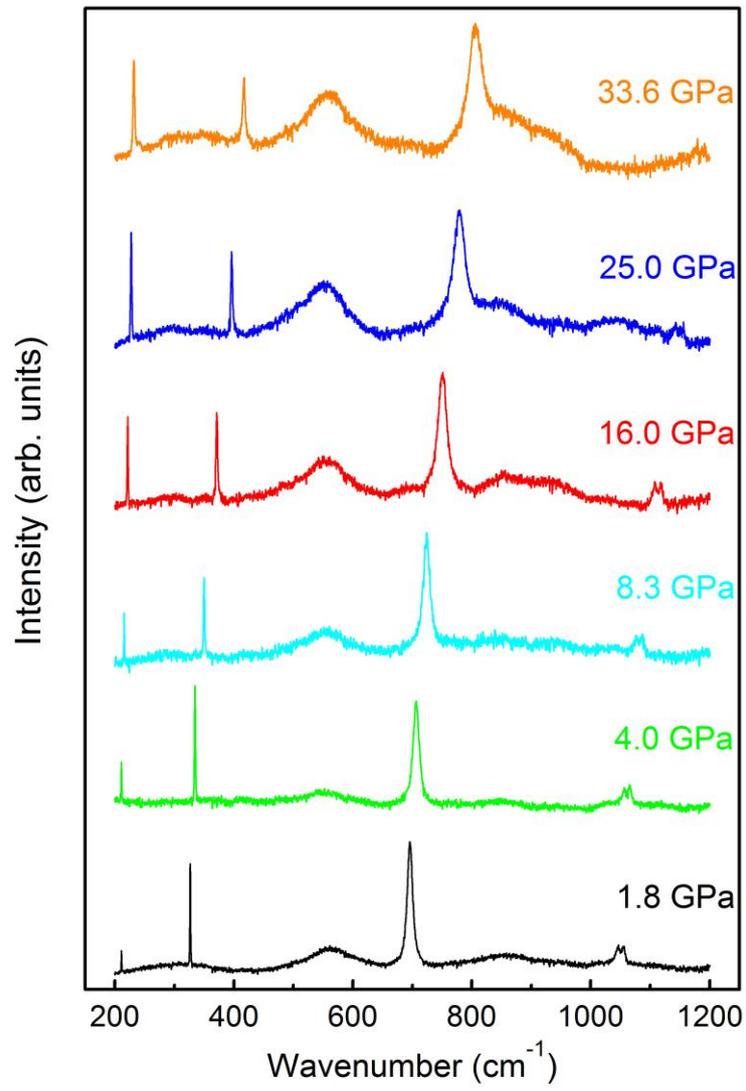

FIG. 8.  Raman spectra of *r*-BS under pressure at room temperature.

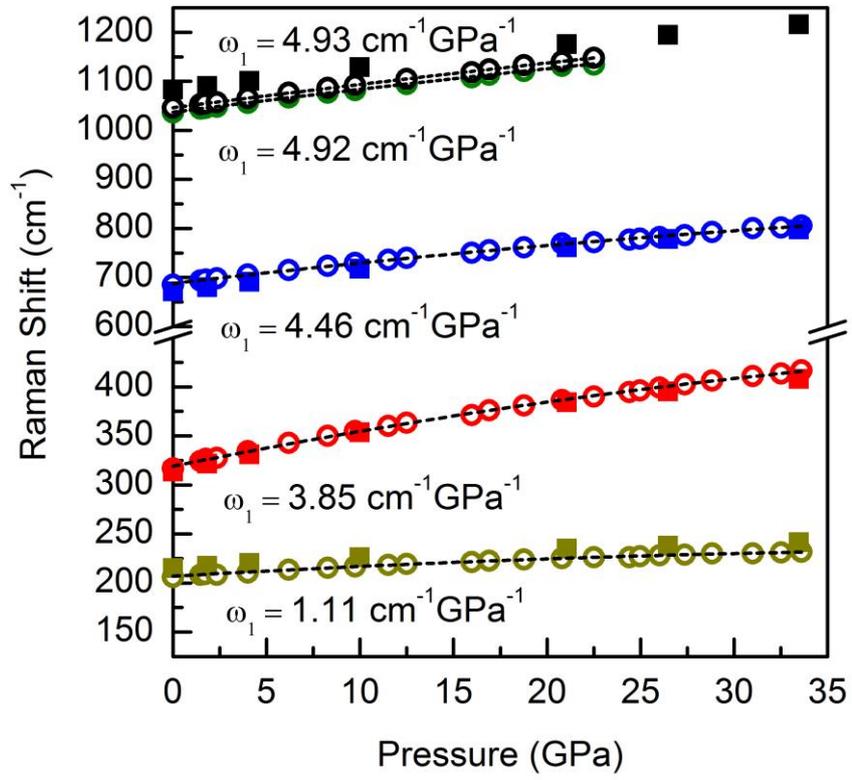

FIG. 9. Pressure dependencies of the phonon mode frequencies experimentally observed (open circles) and theoretically calculated (solid squares). Dashed lines are quadratic least squares fits ($R^2 > 0.999$); the $\omega_1$ parameters of Eq. (5) are listed.